\documentclass{article}
\usepackage{spconf,amsmath,graphicx}
\usepackage{url}
\usepackage[dvipdfmx]{hyperref}
\newcommand{\vm}[1]{\mbox{\boldmath $#1$}}

\newcommand{\argmin}{\mathop{\rm arg~min}\limits}

\usepackage{setspace}
\usepackage{multicol}

\title{crank: an open-source software for nonparallel voice conversion based on vector-quantized variational autoencoder}
\name{
    \begin{tabular}{c}
    Kazuhiro Kobayashi$^{1,2}$,
    Wen-Chin Huang$^{1}$,
    Yi-Chiao Wu$^{1}$, 
    Patrick Lumban Tobing$^{1,2}$, \\
    Tomoki Hayashi$^{1,2}$,
    Tomoki Toda$^{1}$
    \end{tabular}
}
\address{
$^1$Nagoya University, Japan, \\
$^2$TARVO, Inc., Japan, \\
kazuhiro.kobayashi@g.sp.m.is.nagoya-u.ac.jp
}
\begin{document}
%
\maketitle
%
\begin{abstract}
In this paper, we present an open-source software for developing a nonparallel voice conversion (VC) system named crank.
Although we have released an open-source VC software based on the Gaussian mixture model named sprocket in the last VC Challenge, it is not straightforward to apply any speech corpus because it is necessary to prepare parallel utterances of source and target speakers to model a statistical conversion function.
To address this issue, in this study, we developed a new open-source VC software that enables users to model the conversion function by using only a nonparallel speech corpus.
For implementing the VC software, we used a vector-quantized variational autoencoder (VQVAE).
To rapidly examine the effectiveness of recent technologies developed in this research field, crank also supports several representative works for autoencoder-based VC methods such as the use of hierarchical architectures, cyclic architectures, generative adversarial networks, speaker adversarial training, and neural vocoders.
Moreover, it is possible to automatically estimate objective measures such as mel-cepstrum distortion and pseudo mean opinion score based on MOSNet. 
In this paper, we describe representative functions developed in crank and make brief comparisons by objective evaluations. 

\end{abstract}

\section{Introduction}\label{sec:intro}
VC is a technique used to convert paralinguistic information such as gender, speaker individuality, and emotions beyond their physical constraints while keeping the linguistic information of a source speech~\cite{Toda14VC}.
One of main goals in VC research is to freely control arbitrary factors of a source voice into objective factors depending on the situation in which VC is used.
However, control capabilities and the sound quality of the converted voice are usually degraded due to the insufficient modeling accuracy of speech production.
If individual speakers could freely control various factors of the speech, it would open up an entirely new speech communication style.

VC research was initially started to develop a speaker individuality conversion technique enabling a source speaker to change his/her speaker individuality to that of another target speaker while preserving the linguistic content.
In this technique, a statistical mapping function that converts acoustic features of the source speech into those of the target speech is preliminarily trained using a parallel dataset consisting of source and target speakers' utterances with the same linguistic contents.
To improve the modeling accuracy of the statistical mapping function, several techniques such as the use of the Gaussian mixture model (GMM)~\cite{Stylianou98} and deep neural networks~\cite{Sun15} have been proposed.

End-to-end VC~\cite{Huang19,Zhang20} is one of the most powerful mapping techniques using a parallel dataset.
Unlike conventional statistical mapping techniques, it is not necessary to explicitly calculate alignment functions between source and target utterances.
That is, the estimation process of the alignment functions is implicitly included in the model training based on sequence-to-sequence networks and their attention mechanisms.
These techniques usually yield considerable improvements of conversion performance compared with the conventional methods using explicit alignment functions.
Moreover, it is also possible to convert not only the voice timbre but also prosodic information.
On the other hand, it is not straightforward to train the end-to-end mapping function only using a small number of training utterances.
Therefore, collecting many parallel utterances usually becomes a burden for users to develop end-to-end VC systems.

To ease the burden of collecting parallel utterances, nonparallel VC has been developed.
There are two major nonparallel VC techniques, namely phonetic posteriorgram (PPG)-based VC methods~\cite{Sun16,Zhang20} and autoencoder-based VC methods including those using the variational autoencoder (VAE)~\cite{Hsu16,Kameoka19ACVAE,Patrick19,Huang20}, vector-quantized VAE (VQVAE)~\cite{Oord17,Ding19,Van20,Ho20,Haitong20}, and generative adversarial network (GAN)~\cite{Hsu17,kameoka18,Fang18,kaneko19}.
For the PPG-based VC method, the PPG vector is first estimated using a preliminarily trained automatic speech recognition (ASR) system.
Then, the conversion function is trained using the PPG vector and acoustic features of the target speech.
The PPG-based methods achieve relatively higher performance than the autoencoder-based VC methods owing to the speaker-independent linguistic feature of the PPG.
However, it is necessary to prepare many training utterances, including contextual information, to build the ASR system for extracting a convincing PPG vector.  
On the other hand, the autoencoder-based VC methods do not rely on any supervised label such as context labels or parallel utterances, excluding speaker labels.
Therefore, the autoencoder-based VC methods are straightforward for building the VC systems compared with the VC methods using parallel utterances and PPG-based VC methods.

In this paper, we introduce an open-source nonparallel VC software based on VQVAE named crank.
In addition to the VQVAE-based VC method, we have implemented several components such as WaveNet-like~\cite{Oord16wavenet} encoder/decoder networks, the hierarchical architecture, the cyclic architecture, and generative adversarial networks.
Using crank, One may possible to easily 1) reproduce the VQVAE-based nonparallel VC method using preliminarily stored recipes such as Voice Conversion Challenge (VCC) 2018 and VCC 2020, and 2) develop a nonparallel VC system using one's own speech corpus.
In this paper, we describe 1) technical details and usage, 2) brief comparisons with VCC baseline systems, and 3) experimental results of objective measures calculated using the VCC 2018 dataset.

\section{Nonparallel voice conversion based on VQVAE}\label{sec:vqvae}
VQVAE-based voice conversion takes training and conversion phases.

In the training phase, the original feature vector $\vm{x}$ is modeled by the VQVAE consisting of encoder/decoder networks based on the reconstructed loss.
The encoder network encodes the original feature vector into the latent vector $\vm{h}$.
The latent vector is quantized into the discrete latent symbol $\vm{q}$, which minimize the distance between the latent vector $\vm{h}$ and the codebook $\vm{e}$.
\begin{equation}
    \vm{q} = \vm{e}_{k} \;\; \text{where} \;\; k = \argmin_{j} || \vm{h} - \vm{e}_{j} ||_{2}.
\end{equation}
Then, the decoder network generates the reconstructed feature vector $\vm{\hat{x}}$ conditioned on the discrete latent symbol $\vm{q}$ and the auxiliary features $\vm{c}_{org}$ such as the speaker code and $F_0$ of the original speech sample.
The objective function of VQVAE is as follows:
\begin{equation}
    \mathcal{L}_{obj} = || \vm{x} - \vm{\hat{x}} ||^{2}_{2} + || \text{sg}[\vm{h}] - \vm{e} ||^{2}_{2} + \beta || \vm{h} - \text{sg} [\vm{e}] || ^{2}_{2},
\end{equation}
where $|| \vm{x} - \vm{\hat{x}} ||^{2}_{2}$, $|| \text{sg}[\vm{h}] - \vm{e} ||^{2}_{2}$, and $|| \vm{h} - \text{sg} [\vm{e}] || ^{2}_{2}$ are the reconstruction loss, codebook loss, and commitment loss, respectively.
$\beta$ and $\text{sg}[\cdot]$ indicate the hyperparameter for the commitment loss and the stop gradient function, respectively.
Because there is an $\argmin$ function to find the discrete latent symbol, it is not straightforward to optimize this network.
To avoid this problem, VQVAE utilizes a straight-through estimator~\cite{Bengio13} to pass through gradients from the decoder to the encoder via the vector-quantizer.

In the conversion phase, the original feature vector $\vm{x}'$ is first encoded into the latent vector $\vm{h}'$ to find the discrete latent symbol $\vm{q}'$ on the basis of the trained encoder network and codebook.
Then, the target auxiliary features $\vm{c}_{tar}$ with the codebook of predicted discrete latent symbols $\vm{q}'$ are fed into the decoder network to generate a converted feature vector $\vm{\hat{y}'}$.

\section{crank}\label{sec:crank}
crank is an open-source software that implements nonparallel VC frameworks.
The license of crank is linked to the MIT license.
The implementation of crank has been continued on a GitHub repository\footnote{\url{https://github.com/k2kobayashi/crank}}.
In this section, we introduce the basic structure of the crank recipe and representative components.
Table~\ref{tab:comparison} shows the fundamental differences in features between crank and successive VCC baseline systems.

\begin{table*}[!ht]
    \centering
    \caption{Features of crank and successive VCC baseline systems.}
    \label{tab:comparison}
    \begin{tabular}{l|c|c|c|c|c}
        Name                     & Year       & Method & Model & Requirements & Open source \\ \hline
        Baseline~\cite{festvox}                 & 2016       & GMM-based VC & GMM & Parallel data & Yes \\
        sprocket~\cite{sprocket} & 2018       & GMM-based differential VC & GMM & Parallel data & Yes \\
        Merlin~\cite{Wu16}       & 2018       & DNN-based VC& DNN & Parallel data & Yes \\
        Baseline T11             & 2020       & PPG-based VC & LSTM & Speech and context & No \\
        Baseline T16~\cite{Patrick20}  & 2020       & VAE-based VC & VAE& Speech &Yes        \\
        Baseline T22~\cite{Huang20} & 2020      & ASR and TTS & Transformer & Speech and context & Yes \\ \hline
        crank  &  & VQVAE-based VC & VQVAE & Speech     & Yes   \\
    \end{tabular}
\end{table*}

\subsection{Template recipe}
In most open-source software for developing a VC system, it is necessary to prepare a speech dataset to run their fundamental functions.
To rapidly reproduce and confirm the effectiveness of VQVAE-based VC techniques, we have prepared VCC 2018~\cite{VCC18} and VCC 2020~\cite{VCC20} recipes on the basis of the Kaldi recipe~\cite{Kaldi}.
In these recipes, several steps such as downloading the dataset, feature extraction, training, conversion, and evaluation are automatically processed by simply typing a single command after preparing the execution environment for Python3.
Because these recipes are used with a shared template written in shell scripts, it is straightforward to adapt them to one's dataset\footnote{Please see README.md to know how to build an original recipe.}.

\subsection{Feature vector and vocoder}
We supported two kinds of the commonly used feature vector in this research field, namely, the mel-cepstrum parameterized from the spectral envelope extracted by CheapTrick~\cite{Morise15} and the mel-filterbank.
For the VC using mel-cepstrum, it is straightforward to acquire reasonable sound quality using a traditional source-filter vocoder.
On the other hand, for the mel-filterbank, it is not straightforward to achieve acceptable sound quality by using only the Griffin--Lim algorithm~\cite{GriffinLim}.
To avoid this problem, we have integrated the ParallelWaveGAN~\cite{Yamamoto20} vocoder for mel-filterbank decoding\footnote{We used an unofficial implementation of ParallelWaveGAN. \url{https://github.com/kan-bayashi/ParallelWaveGAN}}.

\begin{figure}[!t]
    \centerline{\includegraphics[width=80mm]{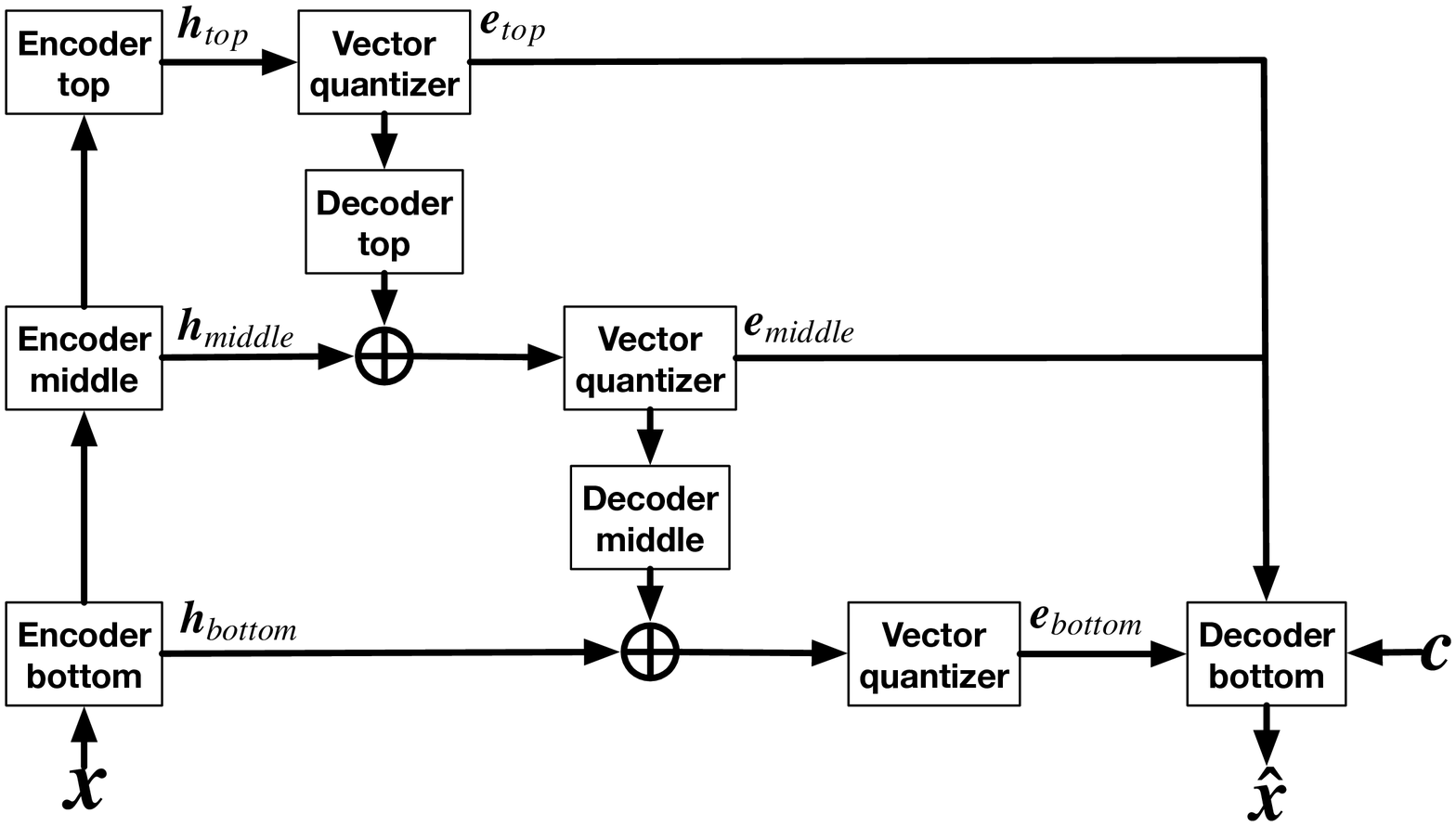}}
    \caption{Architecture of hierarchical VQVAE.}
    \label{hierarchical}
\end{figure}
\subsection{Network architecture}
For the encoder/decoder architecture of the VQVAE network, we used the WaveNet-like network structure~\cite{Oord16wavenet} to achieve higher modeling accuracy and higher inference speed than the recurrent neural network-based network.
The WaveNet-like network structure has several features such as dilated convolution, gated linear unit, and residual connections.
Moreover, as the basis of a work on image generation~\cite{Razavi19}, we used a hierarchical VQVAE structure~\cite{Ho20} to achieve higher modeling accuracy.

Figure~\ref{hierarchical} shows the hierarchical VQVAE structure.
The original feature vector passes three encoders to extract latent vectors in each resolution.
The top vector-quantizer estimates a discrete latent symbol, and then this symbol is fed into the top decoder to adjust the resolution for adding the middle encoder output.
The resulting output is fed into the middle vector-quantizer to estimate the discrete latent symbol, and this latent symbol is also passed through the middle decoder to calculate the discrete latent symbol of the middle stack.
Finally, discrete latent symbols calculated in each stack are concatenated and then fed into the bottom decoder to generate a converted feature vector.
Note that we did not implement down/up-sample functions among time axes in the hierarchical structure to implement a causal network. 

\subsection{Cyclic architecture}\label{subsec:cycle}
A cyclic architecture for nonparallel VC was initially proposed in a VAE-based VC method~\cite{Patrick19}.
On the basis of this work, we implement the cyclic VQVAE-based VC method.
An advantage of the cyclic architecture is that it can include source-to-target conversion flow and target-to-source conversion during its optimization process.
The original feature vector is first converted into the converted feature vector conditioned on auxiliary features for target speaker.
The converted feature vector is converted into the reconstructed feature vector using auxiliary features consisting of source speaker information to calculate reconstruction loss.
By taking into consideration the source-target-source conversion, one can regularize latent features associated with linguistic information among all training speakers.
As a result, it is possible to perform stable speaker individuality conversion for any source-target speaker pairs.

\subsection{Adversarial training}\label{subsec:gan}
The GAN is one of the most powerful frameworks to generate realistic samples.
We implement the least-square GAN framework into the decoder of the VQVAE-based VC.
In the VQVAE-based VC method, the discrete latent symbol is estimated from the vector-quantizer through encoder networks.
By applying the stop gradient function to the discrete latent symbol, the decoder can be regarded as a generator conditioned on a discrete latent symbol and auxiliary features. 
For the discriminator, crank also used a WaveNet-like network structure, and auxiliary classifier GAN~\cite{Odena17} is also implemented to make the training more stable.
Note that it is possible to select either a reconstructed feature vector or a converted feature vector for calculating adversarial loss in our implementation.
On the basis of the ParallelWaveGAN~\cite{Yamamoto20}, we have also implemented a multiresolution short-time Fourier transform (STFT) loss for calculating reconstruction loss.

In VQVAE, a discrete latent symbol is shared over all training speakers as linguistic information.
However, as the training process progresses, the VQVAE network easily suffers from overfitting problems.
As a result, the predicted discrete latent symbol tends to be speaker-specific linguistic information, degrading conversion quality.
To avoid this problem, inspired by the work on ASR~\cite{Shinohara16}, we have implemented an adversarial training procedure for the encoder using a gradient reversal layer. 
To calculate the adversarial loss of the speaker classifier, we also used the same structure as that of the discriminator.

\subsection{Objective measures}
For the nonparallel VC method, overfitting is one of the biggest problems because it cannot directly optimize source-to-target mapping functions.
To avoid this problem, it is reasonable to calculate objective measures that represent conversion performance.
To estimate the conversion performance without performing subjective tests, crank automatically calculates mel-cepstrum distortion and the mean opinion score on the basis of the MOSNet~\cite{lo19} using an evaluation set.
By calculating mel-cepstrum distortion, one can roughly estimate the conversion performance of speaker individuality.
On the basis of MOSNet prediction, it is possible to investigate the sound quality of the converted voice.

\section{Experiments}\label{sec:experiments}
As brief comparisons between representative functions developed using crank, we evaluated objective measures calculated using VCC 2018 recipes.
The sampling rate was set to 22050~Hz.
The number of training speakers was 12, and each speaker spoke 80 utterances.
We used 75 utterances for training and the remaining five utterances for development.
We used the other 35 evaluation utterances in each speaker.
The evaluation was performed under the speaker-closed condition (i.e., training speakers were used for the evaluation as well.).

An 80-dimensional mel-filterbank was used as the feature vector.
Continuous $F_0$, an unvoiced/voice decision symbol, and a speaker code were used as the auxiliary features for the decoder.
The ParalellWaveGAN vocoder trained using the same dataset was used as a neural vocoder.
We compared mel-cepstrum distortion and predicted naturalness on the basis of MOSNet in this evaluation.
The values were averaged among all-speaker pairs.
We used a 35-dimensional mel-cepstrum to calculate the distortion.
The other settings and resulting voices were described on the website\footnote{\url{https://k2kobayashi.github.io/crankSamples/}}.

The following techniques were compared in this evaluation.
\begin{description}
    \item[Baseline VQVAE] \mbox{}\\
        Three-stacked hierarchical VQVAE
    \item[CycleVQVAE] \mbox{}\\
        Baseline VQVAE with cyclic architecture 
    \item[VQVAEGAN] \mbox{}\\
        Baseline VQVAE with GAN
    \item[CycleVQVAEGAN] \mbox{}\\
        Baseline VQVAE with cyclic architecture and GAN
    \item[CycleVQVAEGAN w/ STFTLoss] \mbox{}\\
        Baseline VQVAE with cyclic architecture and GAN with STFT loss
\end{description}
The STFT loss means that the network utilizes not only L1 loss but also STFT loss for the reconstruction loss.

Table~\ref{tab:objective} shows the experimental results of the mel-cepstrum distortion and MOSNet predictions.
Compared with the baseline VQVAE, the CycleVQVAE method achieves higher performance in terms of Mel-CD and MOSNet.
Moreover, integrating GAN-based training, we can see that the CycleVQVAE w/ STFT method yields the highest performance among methods shown in Table~\ref{tab:objective}.
On the other hand, the VQVAEGAN method has a lower performance than the baseline VQVAE method.
It is considered that it is not straightforward to optimize the VQVAE decoder network based on the GAN framework, and a cyclic architecture may maintain the stability of the training similarly to CycleGAN-VC~\cite{kaneko19}.

\begin{table}[]
    \centering
    \caption{Objective evaluations.}
    \label{tab:objective}
    \begin{tabular}{l|c|c}
        Method                    & Mel-CD      & MOSNet        \\ \hline
        Baseline VQVAE            & 9.89        & 3.53          \\
        CycleVQVAE                & 9.66        & 3.54          \\
        VQVAEGAN                  & 10.13       & 3.44          \\
        CycleVQVAEGAN             & 9.74        & 3.48          \\
        CycleVQVAEGAN w/ STFTLoss & \bf 9.64    & \bf 3.59      \\
    \end{tabular}
\end{table}

\section{Conclusion}\label{sec:conclusion}
In this paper, we introduced an open-source nonparallel VC software named crank.
The main objective of developing crank is to build a nonparallel VC system with limited constraints for collecting the speech corpus.
In addition to the vector-quantized variational autoencoder-based VC method, several representative methods such as these using the hierarchical architecture, cyclic architecture, generative adversarial network, and speaker adversarial training have been implemented in crank. 
Moreover, it also supports the ParallelWaveGAN vocoder to decode a converted mel-filterbank and calculate objective measures such as mel-cepstrum distortion and pseudo mean opinion score on the basis of MOSNet.
For our future work, we will continue to develop methods to realize high-quality, easy-to-use nonparallel VC software.

\section*{Acknowledgment}
This work was partly supported by JSPS KAKENHI Grant-in-Aid for JSPS Research Fellow Number 19K20295, and JST, CREST Grant Number JPMJCR19A3.


\ninept
\bibliographystyle{IEEEbib}
\bibliography{bib/vc,bib/url,bib/nn,bib/asr}

\end{document}